# Comparative analysis of taper models for *Pinus nigra:* A study across parametric, semi-parametric, and non-parametric models using terrestrial laser scanner acquired data

I. Boukhris[1,2*], N. Puletti[3T], C. Vonderach[4,5T], M. Guasti[3], S. Lahssini[6], M. Santini[7], R. Valentini[1,2]

1. Department for Innovation in Biological, Agri-Food and Forest Systems (DIBAF), University of Tuscia, 01100 Viterbo, Italy (issam.boukhris@unitus.it; rik@unitus.it)

2. Division Impacts on Agriculture, Forests and Ecosystem Services (IAFES), Fondazione Centro Euro-Mediterraneo sui Cambiamenti Climatici, 01100 Viterbo, Italy (monia.santini@cmcc.it)

3. CREA, Research Centre for Forestry and Wood, Viale Santa Margherita 80, IT-52100 Arezzo, Italy (nicola.puletti@crea.gov.it; matteo.guasti@crea.gov.it)

4. Forest Research Institute Baden-Württemberg, 79100, Freiburg, Germany (christian.vonderach@forst.bwl.de)

5. Chair of Forest Growth and Dendroecology, University of Freiburg, Tennenbacherstaße 4, 79106 Freiburg, Germany

6. Department of Forest Development, National School of Forest Engineers, 11000 Salé, Morocco (marghadi@gmail.com)

*issam.boukhris@unitus.it

[T]Authors contributed equally to this work

**Abstract**

**Key message**

Taper equations are essential tools for characterizing tree stem profiles, offering valuable insights for forest management, timber inventory, and assortments allocation. Developing taper models requires multiple observations on upper-stem diameters and other covariates collected on standing or felled trees. Recent advancements in Terrestrial Laser Scanning technology have revolutionized data acquisition. It provides a remarkable opportunity for precise, non-invasive data collection while saving time compared to traditional methods. Beyond the data collection method, the thoughtful selection of the appropriate model category and form is paramount. This decision significantly shapes the accuracy of predictions, making it a crucial consideration in the taper modeling process.

**Context**

The careful choice of an appropriate taper model form is crucial to ensure the accuracy of predictions. Therefore, a comprehensive investigation into the performance of models from diverse class categories, fitted to data extracted from Terrestrial Laser Scanner point clouds, is a prerequisite before establishing a model for its various applications.

**Aims**

The aims were to evaluate the performance of four taper models from three different model categories fitted to data extracted from Terrestrial Laser Scanner point clouds in predicting diameter over bark (dob) and total stem volume, specifically within *Pinus nigra* stands in the Vallombrosa forest, situated in the North Apennine mountains of Italy.

**Methods**

Four taper models, representing three distinct model categories (parametric, semi-parametric, and non-parametric), were established based on point cloud data collected from 219 *Pinus nigra* trees. The performance of these models in predicting both dob and total stem volume was rigorously assessed using K-fold cross-validation, which involved the evaluation of key metrics, including the squared estimate of errors (SSE), bias (B), and fitness index (FI). Additionally, the models' ability to predict diameter across various relative height classes and volume within different diameter at breast height (DBH) categories was thoroughly examined, and the models were subsequently ranked based on these criteria.

**Results**

The results show that among fitted models, the Max and Burkhart segmented model calibrated by the means of a mixed-effects approach provided the best estimate of the diameter at different heights and the total stem volume evaluated for different diameter at breast height (DBH) classes. In numerical terms, this model estimated the diameter and the volume with a respective overall error of 0.781 cm and 0.021 $m^3$. The predicted profile also shows that above a relative height of 0.7, the diameter error tends to increase due to the low reliability of data collected beyond the base of the crown primarily caused by interference from branches and leaves.

**Conclusion**

Accurate taper model selection is fundamental for forest management. The Max and Burkhart segmented model, calibrated with a mixed-effects approach, demonstrated the best accuracy. However, data collection above a relative height of 0.7 faced challenges. Nevertheless, Terrestrial Laser Scanning (TLS) holds promise as a non-destructive alternative for taper profiles and tree volume estimation.

## Introduction

Precision forestry is becoming increasingly important in the face of the urgent challenges posed by climate change and the rising call for sustainable forest management (SFM) practices (Kovácsová & Antalová, 2010). This new emerging direction relies on the use of accurate data acquired by means of advanced technologies to make informed decisions (Fardusi et al., 2017). In the realm of SFM, a noteworthy potential exists for value creation through implementing improved practices. Besides the ecological advantages related to enhanced productivity, which contribute to carbon sequestration and alleviate pressure on forests, the adoption of this paradigm shift is fundamentally linked with substantial economic and social value (Choudhry & O'Kelly, 2018).

One concrete and increasingly prevalent instance of the potential being realized is precision harvesting (PH), which aims to maximize the efficiency of the harvesting operation while minimizing the impact on the environment and overcoming several limitations inherent to conventional logging practices (e.g., lack of precision and inefficient resource utilization due to lack of tree-level data) (Olivera Farias & Visser, 2016). Within the scope of PH, the optimal allocation of assortments is a crucial component of the wood products supply chain and the carbon stock projections since different wood products hold different economical values and carbon storage potential (Brunet-Navarro et al., 2017; Marchi et al., 2020). Stem taper equations, which predict the change in stem form from ground to tip are fundamental tools to accurately disaggregate trees into specific products based on certain specifications like log length and diameter and can help reach a better optimization of wood products (Calders et al., 2015, 2022; Puletti et al., 2019).

In the academic literature, a panoply of taper model formulations and methods of parameter estimation is currently in use and the selection of a proper model form is more important than the actual fitting method (Weiskittel et al., 2011). Following the classification of McTague & Weiskittel (2020), taper models could be grouped into three main categories: **i-** parametric models (e.g., simple-taper, variable-exponent, and segmented equations) which include most taper equations, are calibrated by the means of parametric approaches (e.g., nonlinear least squares (NLS) and nonlinear mixed effects (NLME) methods) and can ensure biologically consistent behavior (Mäkelä & Valentine, 2020) **ii-** semiparametric models (e.g., B-splines and P-splines) which offer greater flexibility in the fit without requiring an extensive addition of parameters, may provide a better representation for complex form species (Kuželka & Marušák, 2014) **iii-** nonparametric models (e.g., Random Forest (RF) and Artificial Neural Network (ANN)) can offer a strong predictive performance without requiring the testing of statistical hypotheses (Ruth E. Baker et al., 2018).

Developing taper equations requires multiple observations on upper-stem diameters and other covariates collected on standing or felled trees (Gómez-García et al., 2013). The recent progress in Terrestrial Laser Scanning (TLS) technology provides an unprecedented prospect for acquiring necessary data with high accuracy and non-destructive means while reducing the required time compared to conventional methods (Calders et al., 2015; Pitkänen et al., 2019). Data acquisition for TLS-based field measurements can be performed following single or multi-scan approaches. The single-scan approach has the simplest data acquisition setting and the fastest speed, however, a major problem with that scan method is that only parts of the trees are covered due to occlusion effects by other objects. On the other hand, the multi-scan approach appears the most accurate technique and has the potential to fully cover trees within the sampling area since scans are performed from multiple directions to overcome the occlusion issue (Liang et al., 2016). To illustrate, many studies using the multi-scan mode reported a stem detection rate between 62.1% and 100% depending on the forest structure and the scanning setup (Liang & Hyyppä, 2013; Maas et al., 2008).

*Pinus nigra* is a fast-growing conifer with a wide, but fragmented distribution across Europe and Asia Minor, predominantly in mountainous areas. Due to its ecological flexibility, it is one of the most widely used tree species for reforestation worldwide, and it is considered a potential substitute for native coniferous species in Central Europe under future climate change (Thiel et al., 2012). Economically, it is one of the most important conifers in Southern Europe, as its wood is highly suitable for general construction, indoor flooring, furniture industry, fuelwood, and paper pulp (Enescu et al., 2016). In Italy, *Pinus nigra* stands cover an area of 444 785 hectares, representing 4.25% of the total forest area, with an age range from 50 to 95 years (Marchi et al., 2020). Taking the Vallombrosa forest in the northern Apennine mountains of Italy as an example, *Pinus nigra* stands occupy 11% of the total area of this forest and are considered an important resource for wood industries in the surrounding region (Bottalico et al., 2012). The previously developed yield equations for *Pinus nigra* were based on data collected from young trees and have been shown to underestimate the volume of mature trees in the current conditions. As such, the primary aim of this study was to develop a tool specific to *Pinus nigra* species that provides forest managers with more accurate estimates of volume and enables better optimization of assortments for industrial purposes. To achieve this goal, we collected data using a multi-scan approach with TLS and evaluated the performance of four taper models from the three different model categories in predicting diameter over bark (dob) and total stem volume specifically.

## Study area and data description

### Study area and sampling protocol

The study was conducted in the Vallombrosa forest (43° 44' N, 11° 34' E), a biogenetic reserve located about 50 km east-southeast of Florence, Tuscany, Italy. The native vegetation of this forest is mainly represented by beech (*Fagus sylvatica*) at higher elevations, oak-hornbeam stands (Quercus spp. mixed with Carpinus betulus L. and Ostrya carpinifolia Scop.), and chestnut (Castanea sativa Mill.) at lower altitudes (Dálya et al., 2019). As part of the main reforestation program that occurred after the World War II, the European black pine (Pinus nigra Arn.) was introduced in the extreme western side of this forest (Bottalico et al., 2012), which was specifically studied here.

The *Pinus nigra* area was stratified as part of the sampling scheme based on two criteria: stand density and total tree height. Each criterion was divided into three levels to ensure optimal representation of forest stand variability, resulting in a total of nine strata. An airborne lidar flight was utilized to achieve this stratification, providing high-resolution data on forest structure and height. Subsequently, one scan center was randomly placed in each stratum, then, a multi-scan TLS mission was performed on each scan center to collect detailed information on forest characteristics (Figure 1).

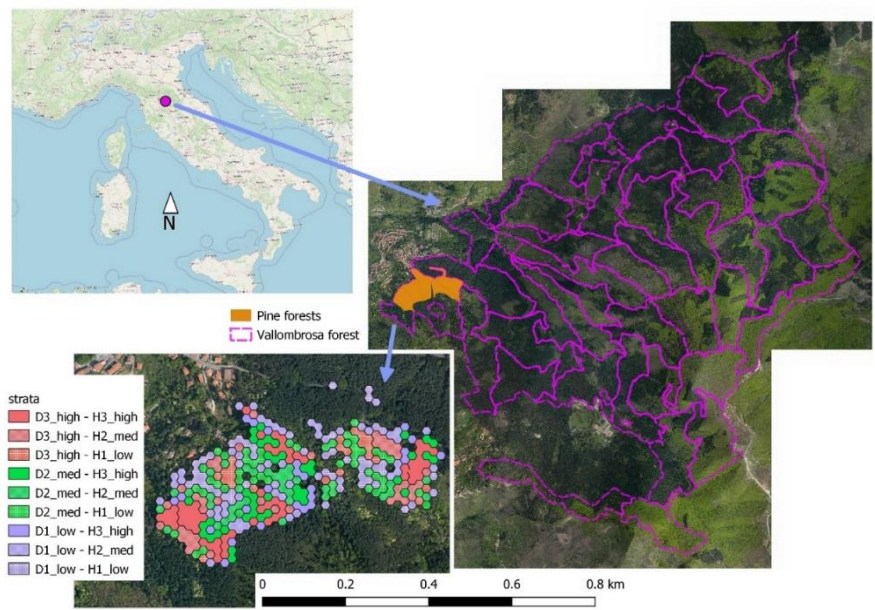

Figure 1: Map of the geographical situation of the Vallombrosa forest and the location of *Pinus nigra* stands in this forest (Orange color). Stratification of those stands by density (D1-D3) and canopy height (H1-H3; down left sub-figure).

### Data collection with TLS

TLS data were acquired using a FARO Focus 3D x 130 (FARO Technologies Inc., Lake Mary, FL, USA). The instrument uses a phase-shift-based technology with a maximum range of 120 m and acquires data with an azimuth scan angle of 360°. It collects the x, y, and z coordinates, and the intensity of the laser returns with a scan-ranging noise of ±1 mm.

Using the TLS device, the data acquisition was set to a reasonable scan configuration providing a good trade-off between a sufficient density and the required time for a single scan (resolution of 7.6 mm at 10 m and 1/5 4x overall quality), for a total of about 28 million pulses per scan. With the aim of reducing occlusion due to other obstacles or vegetation and ensuring sufficient coverage, scanning was performed in multiple positions, at least 8 scans per plot subjectively distributed based on the density and structure. Each of the trees in the stand was at least scanned from 3 positions and up to 12 white polystyrene registration spheres (14 cm diameter) were placed in each plot to aid in the digital registration of individual scans.

**Data processing**

Individual scans were merged on the plot level using the automatic registration algorithm implemented in Trimble Real Works® (TRW) software. The program automatically joins overlapping redundant points to create one seamless 3D point cloud suited for the analysis. Details of the operation (settings, criteria, and thresholds) performed by the software are not declared nor accessible. The process attained a very low plot-level registration error, and TRW achieved a high precision scan placement (mean = 2.33 mm, sd = 3.18 mm, max = 8.21 mm). Using the TRW tool, each pine tree stem was manually separated from the rest (i.e., from the ground, the crown, and other non-pine points). Diameter measurements were acquired at approximately 1-meter intervals through the fitting of cylinders to point cloud data. To determine the volume, Huber's formula was employed, which entails utilizing the diameter measured at the midpoint of a log segment (Figure 2).

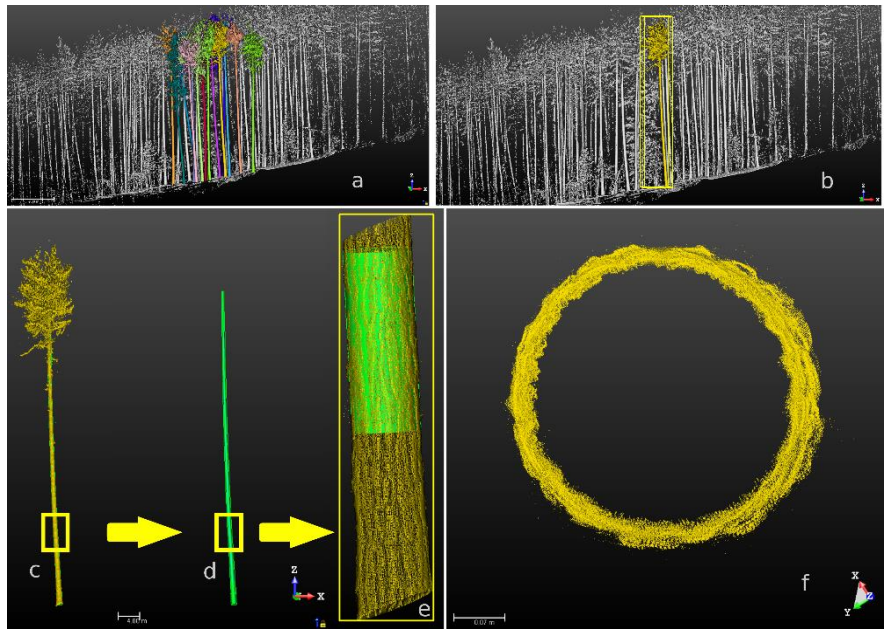

Figure 2: Schematic illustration of the point cloud data exploration. a- The resulting point cloud after the pre-processing steps from a small region within one stratification unit; b- Identification of a representative tree within the point cloud; c- Zoomed-in view of the stem of the identified tree; d-

Removal of branches for stem cleaning; e- Fitting a cylinder to a specific section of the tree; f- Cross-sectional view of the point cloud for the section depicted in subfigure-e.

**Data presentation**

Detailed descriptive statistics are provided for the trees inventoried in this study (Table 1). Statistics are derived from stem attributes estimated from point cloud data collected by the TLS device. The diameters and total height of the stems were estimated using the method described earlier, while the stem volume was determined by applying the Huber formula to each of the 1m-length cylinders. In total, 219 trees were extracted from the inventoried area and were systematically allocated to 5 folds. Four folds were exclusively designated for model training, while the remaining fold was reserved for rigorous testing during each iteration of the cross-validation procedure.

Table 1: Mean, standard deviation, and range for tree characteristics from all the samples.

|  | **Mean** | **SD** | **min** | **max** |
|---|---|---|---|---|
| **Full dataset (N = 219)** | | | | |
| DBH (cm) | 45.05 | 11.22 | 21.0 | 82.0 |
| Total height (m) | 30.7 | 4.0 | 20.5 | 40.5 |
| Disk* dob (cm) | 29.3 | 12.5 | 0 | 82.5 |
| Disk height (m) | 14.5 | 8.7 | 0.9 | 40.5 |
| Crown base height (m) | 21.0 | 4.0 | 11.7 | 31.3 |
| Volume (m$^3$) | 2.27 | 1.32 | 0.38 | 8.04 |

*Disk refers to a given cross-section of the stem for which the diameter is measured

**Materials and Methods**

The data acquired by TLS was used to fit four taper models from three different model categories, namely parametric, semi-parametric, and non-parametric. Thereafter, the fitted models were evaluated in terms of their ability to predict diameters at different heights as well as the total stem volume. In the following, the different used taper models, as well as the evaluation metrics, are explicitly introduced.

**Max and Burkhart (1976)**

The segmented stem profile model developed by (Max & Burkhart, 1976) was used to represent the parametric model's category. Numerous studies have shown that this model is accurate in estimating upper stem diameters of *Pinus nigra* (e.g., Özçelik & Brooks, 2012). The model consists of three quadratic functions representing the stem's lower, middle, and upper parts grafted together at two points (a1 and a2) with continuous polynomials and continuous first derivatives at each point. In the present study the taper equation was constrained to pass through diameter at breast height (dbh) while it is by construction constrained for the total height (ht).

Max and Burkhart's (1976) taper equation can be expressed as follows:

$$y_{ij}(z_{ij}) = b_1 z_{ij} + b_2 z_{ij}^2 + b_3 (z_{ij} - a_1)^2 I_1 + b_4 (z_{ij} - a_2)^2 I_2 + \varepsilon_{ij} \qquad (1)$$

where $y_{ij} = \left(\frac{d_{ij}}{dbh_i}\right)^2$; $dbh_i$ = dbh of tree i, (i = 1, 2, …. , N); N = number of trees in the sample; $d_{ij}$ = the i$^{th}$ upper-stem diameter over bark (dob) at height $h_{ij}$ on tree i; (j = 1, 2, …., $n_i$); $n_i$ = number of diameter measurements for tree i, $z_{ij} = 1 - \frac{h_{ij}}{ht}$ = the complement of the relative height, $a_1$ and $a_2$ =

join points to be estimated from the data, $I_k = 1$ if $z > a_i$, and 0 otherwise, $k = 1,2$, $b_p$'s = regression coefficients with, p = 1, 2, 3, 4, $\varepsilon_{ij}$ = residual error term associated to the prediction of $y_{ij}$.

To constrain the taper equation to pass through dbh at the breast height, only one regression coefficient needed to be changed. In this case, the $b_1$ parameter was modified as suggested in (Cao, 2009) and replaced by the modified b1* which can be expressed as follows:

$$b1* = \frac{1 - y_i(z_{bh}) + b1*z_{bh}}{z_{bh}} \quad ; \quad z_{bh} = 1 - \frac{1.3}{ht} \tag{2}$$

To estimate the parameters of Eq. 1, the Levenberg-Marquardt algorithm, which is implemented within the nls function in the R stats package (R Core Team, 2022), was utilized.

**Mixed-effects model**

The stem taper models can have a high degree of multicollinearity, autocorrelation, and heteroskedasticity, all of which can be addressed with appropriate statistical methods. Many studies reported a reduction in the within-tree autocorrelation when including a tree-level random effect (e.g., (Cao & Wang, 2011; Garber & Maguire, 2003). Using a mixed model for example, the autocorrelation within the stem can be accounted for at each height measurement level.

In the mixed effect formulation, all parameters of Eq.1 are expressed as fixed effect (population level coefficients) while some parameters may contain additional random effect (for each tree within the population). In the case of the (Max & Burkhart, 1976) taper equation, different combinations of the parameters (b1, b2, b3, b4, a1, a2) could be a potential candidate for random parameters.

Here is an expression for the stem profile including the fixed and random effects:

$$y_{ij}(z_{ij}) = b_{1i}z_{ij} + b_{2i}z_{ij}^2 + b_{3i}(z_{ij} - a_{1i})^2 I_1 + b_4(z_{ij} - a_{2i})^2 I_2 + \varepsilon_{ij} \tag{3}$$

where $a_{1i}$ and $a_{2i}$ = join points to be estimated from the data for each tree i, $I_k = 1$ if $z > a_{ki}$, and 0 otherwise, $k = 1,2$, $b_{pi}$'s = regression coefficients to be estimated for each tree i; p = 1, 2, 3, 4

To define the statistical properties of the terms in Eq. 3, its vectorial formulation presented in the following offers a more simplified notation and will be considered:

$$Y_i = (A_iB + B_iU_i)X_i + \varepsilon_i \tag{4}$$

where $Y_i$ is a vector of predictions, $X_i$ is a matrix of predictors, $A_i$ is a matrix for the fixed effects, $B_i$ is a matrix for random effects, B is a vector of fixed parameters, $U_i$ is the vector for random effects and $\varepsilon_i$ is a vector of random errors, with the assumptions that $\varepsilon_i \sim N(0, R)$ and $U_i \sim N(0, D)$ where R is the diagonal variance-covariance matrix of $\varepsilon_i$ and D is a diagonal variance-covariance matrix of $U_i$.

Using the R saemix package (Comets et al., 2017), the stochastic approximation expectation maximization algorithm was used to estimate the parameters of Eq.3.

The volume equation derived from the integration of the Max and Burkhart taper equation can be expressed as follows:

$$V = K \times dbh^2 \times ht \left[ \frac{b1}{2}(z_y^2 - z_x^2) + \frac{b2}{3}(z_y^3 - z_x^3) + \frac{b3}{3}\left[(z_y - a_1)^3 I_1 - (z_x - a_1)^3 J_1\right] + \frac{b4}{3}\left[(z_y - a_2)^3 I_2 - (z_x - a_2)^3 J_2\right]\right] \tag{5}$$

where $z_y = 1 - \frac{h_y}{ht}$ and $z_x = 1 - \frac{h_x}{ht}$; $h_x$ and $h_y$ being respectively the lower height of interest (m) and the upper height of interest, K = conversion factor = $\frac{\pi}{40000}$, $I_k = 1$ if $z_y > a_i$, and 0 otherwise, $J_k = 1$ if $z_x > a_i$, and 0 otherwise, k=1, 2.

**B-splines**

Semiparametric models are a class of statistical models that provide a flexible compromise between the rigidity of fully parametric models and the complexity of fully nonparametric models. By avoiding rigid assumptions about functional form, semi-parametric models offer the potential to capture the underlying relationship between variables more accurately, particularly for unknown or complex functional forms, while parametric approaches can be used for estimating model coefficients. Basis-splines (B-splines) are a type of semiparametric model that possess favorable numerical properties. To construct a B-spline model, a fixed number of splines, also called polynomials, are connected at specific points known as knots while ensuring the condition of continuity of their second derivatives across the knots.

Here, we follow Kublin et al. 2013 and consider the population mean diameter d for a given relative height (hr) may be approximated by a B-spline function of degree p and can be expressed as follows:

$$d_i = f(hr_i) + \varepsilon = \sum_{l=0}^{d1} \beta_l B_{l,p}^{(1)}(hr_i) + \varepsilon \qquad (6)$$

where $f(hr_i)$ is the response of the model which renders the diameter value $d_i$ at a given $hr_i$, $\beta_l$'s are the parameters of the model, $d_1$ is the number of parameters in the B-spline function with the condition that $d_1 \leq k_1 + 1 + p$; $k_1$ corresponds to the number of internal knots, $B_{l,p}^{(1)}(hr)$ is the B-spline basis function computed based on the "*de Boor recurrence relation*".

Due to the nonlinear nature of stem tapering and the hierarchy of data acquired from sampled trees, semiparametric models can be calibrated more efficiently using mixed approaches, which can also consider the random deviation of an individual tree from the population average. Linear mixed effect models can be used for the calibration if knot values are fixed a priori, which has the advantage of being less computationally expensive and more numerically stable than nonlinear mixed models.

The B-spline based taper equation considering both fixed and random effects can be written as:

$$d_{ij} = f(hr_{ij}) + g(hr_{ij}) + \varepsilon_{ij} = \sum_{l=0}^{d1} \beta_l B_{l,p}^{(1)}(hr_{ij}) + \sum_{l=0}^{d2} \theta_{i,l} B_{l,p}^{(2)}(hr_{ij}) + \varepsilon_{ij} \qquad (7)$$

where $f(hr_{ij})$ corresponds to the population mean response, $g(hr_{ij})$ represents the tree-specific deviation from $f(hr_{ij})$, $\varepsilon_{ij}$ is the tree-specific residual error, $d_{ij}$ corresponds to the j[th] diameter from the i[th] sample tree, $hr_{ij}$ corresponds to the j[th] relative height from the i[th] sample tree, $\beta_l$'s are the fixed regression coefficient of the model, $\theta_{i,l}$'s are the random effect parameters to be estimated for each tree, d2 is the number of random parameters in the g function with the condition that $d_2 \leq k_2 + 1 + p$; $k_2$ corresponds to the number of internal knots in the random effect part of the model, $B_{l,p}^{(2)}(hr_{ij})$ is the B-spline basis function for the random effect part of the model.

The vectorial formulation of Eq. 7 can be expressed as follows:

$$Y_i = X_i \beta + Z_i \theta_i + \varepsilon_i \qquad (8)$$

where $Y_i$ is the vector of diameter predictions at the level of the i[th] tree, $X_i$ is the matrix of B-spline basis values for the fixed effect part of the model, $Z_i$ is the matrix of B-spine basis values for the random effect part of the model, $\beta$ is a vector of fixed parameters, $\theta_i$ is a vector of random effect parameters, $\varepsilon_i$ is the vector of residual error with the assumption that $\varepsilon_i \sim N(0, R)$ and $\theta_i \sim N(0, G)$ where R is the diagonal variance-covariance matrix of $\varepsilon_i$ and G is a positive non-diagonal matrix of $\theta_i$. For more details on this method see (Kublin et al., 2013).

According to the suggestions by (Kublin et al., 2013), the parameters of Eq. 7 were estimated considering knots [0, 0.1, 0.75, 1.0] corresponding to the B-spline basis function $B^{(1)}$(hr) for the population mean function and [0, 0.1, 1.0] for $B^{(2)}$(hr) representing the deviation from the population average. Hence, the regression model depends only on relative height. Measurements of dbh and further diameter measurements of a specific tree are merely used to calibrate the tree specific deviation from the average taper curve by estimating the corresponding random effects. To ensure diameter estimates of zero at tree top, the last spline in both $B^{(1)}$(hr) and $B^{(2)}$(hr) are omitted (see Figure A.1 for a graphical visualization and Kublin et al. 2013).

The R package TapeR was used to calibrate the model by the Restricted Maximum Likelihood (REML) method (default in the internally used R package nlme (Pinheiro et al., 2023)) providing diameter andrelative height values for each sampling tree of training data. To estimate stem volumes, the implemented numerical integration method included in the TapeR package was used (Kublin et al., 2023).

**Random Forest**

Nonparametric methods have been recently used for taper predictions providing reasonable performances (e.g., Nunes & Görgens, 2016; Yang & Burkhart, 2020). The present study used random forest to estimate the diameter at different heights of the stem. This algorithm works by training many decision trees on random subsets of the features, then averaging out their predictions. Regarding the structure of the model, the diameter was set as the dependent variable, which were predicted by four independent predictors namely dbh, h, ht and h/ht (abbreviations as above). For fine-tuning the model, the grid search method was used to evaluate all possible combinations of hyperparameters (viz., the number of decision trees; the number of features to consider when looking for the best split; the minimum number of samples required to split an internal node) within the search space as well as the cross-validation method to prevent overfitting. The Python library scikit-learn (version 0.24.2) which includes an implementation of the Random Forest algorithm was used here (Pedregosa et al., 2011).

Since nonparametric methods are not algebraically integrable, including the random forest algorithm, the Monte Carlo integration, which is a numerical integration approach, was self-implemented and used to predict the stem volume for each sample tree. This technique relies on a random choice of points at which the integrand is evaluated. Monte Carlo integration can be expressed as follows:

$$V = K \times (h_b - h_a) \times \frac{1}{N} \times \sum_1^N (RF(h_i))^2 \approx K \times \int_{h_a}^{h_b} (RF(h))^2 dh \qquad (9)$$

where $h_a$ and $h_b$ being respectively the lower and the upper heights of interest (m), K = conversion factor = $\frac{\pi}{40000}$, $h_i$ corresponds to the i$^{th}$ height on the tree generated randomly from a uniform distribution U($h_a, h_b$); (i = 1, 2, …., N); N is the number of samples generated with the distribution U and was set to 400 following a convergence analysis.

Figure 3 below provides a visual summary of the diameter calculation models used in this study.

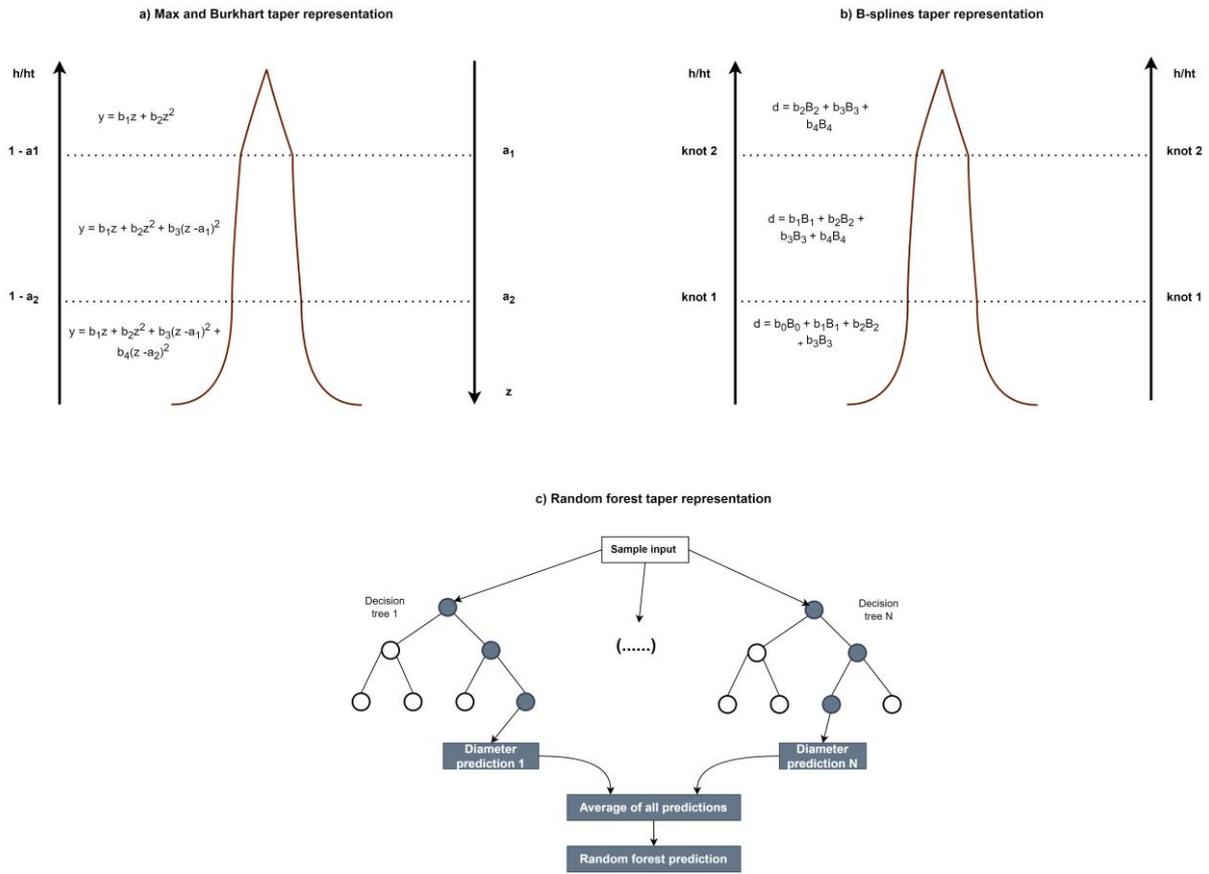

Figure 3: Schematic tree form representation for the three applied models. a) the Max and Burkhart parametric model; b) B-splines semi-parametric model; c) Random forest non-parametric model.

**Evaluation of the methods**

Taper models should yield unbiased estimates with minimal variance in both diameter outside bark (dob) and stem volume. A single overall measure of bias or residual error of estimation calculated for dob and stem volume does not provide an exclusive indicator of the goodness in evaluating several taper equations for a set of trees of a given species. This holds true when calculating bias metrics at different parts of stems and for different trees can result in a zero or close-to-zero average given that bias can be positive or negative. Alternatively, the standard error which can report the variability of the biases is considered of better reliability. Moreover, when dealing with small sample size, the residual variance can be influenced by the degrees of freedom, making the correlation index, also known as the fitness index, a useful metric to address this concern. All those metrics were used to rank the fitted models.

The most reliable measure of mean bias (B), overall residual variance (SSE) and the fitness index (FI) can be expressed as follows:

$$B = \frac{\sum_{i=1}^{n}(\hat{Y}_i - Y_i)}{n} \tag{10}$$

$$SEE = \frac{\sum_{1}^{n}(Y_i - \hat{Y}_i)^2}{n-k} \tag{11}$$

$$FI = 1 - \frac{\sum_{1}^{n}(Y_i - \hat{Y}_i)^2}{\sum_{i=1}^{n}(Y_i - \bar{Y}_i)^2} \tag{12}$$

where $Y_i$ is the actual observation, $\hat{Y}_i$ is the predicted value of the actual observation, $\bar{Y}_i$ is the mean of the actual observations, n is the number of observation and k corresponds to the number of estimated parameters.

As part of our evaluation process, we employed a grouping methodology to assess the performance of the taper models in conjunction with the overall assessment presented earlier (5-fold cross validation). For dob, biases and standard errors were calculated for each 10% step of the height along the stem, while for the total stem volume (from ground to top), the biases and standard errors were calculated by DBH classes. This allowed us to gain a comprehensive understanding of the accuracy of the taper equations across various tree dimensions as well as across different parts within the trees.

## Results

### Taper prediction

The model by Max and Burkhart (1976) was fit using both fixed and mixed-effects approaches. Different mixed-effects combinations of parameters were tested (Table A.1), and the combination involving a1, a2, b1, b2 and b4 produced the lowest values of Akaike's information criterion (AIC) and Bayesian information criterion (BIC), resulting in the following parameter values (Table 2) and model form (Equation 13):

$$y_{ij}(z_{ij}) = b_{1i}z_{ij} + b_{2i}z_{ij}^2 + b_3(z_{ij} - a_{1i})^2 I_1 + b_{4i}(z_{ij} - a_{2i})^2 I_2 + \varepsilon_{ij} \qquad (13)$$

Table 2: Estimates of parameters (and standard errors) for fixed-effects and mixed-effects taper models, based on the validation data set.

| Parameters | Fixed-effect model | Mixed-effects model |
|---|---|---|
| a1 | 0.6384  (0.0242) | 0.71   (0.007) |
| a2 | 0.8756  (0.0043) | 0.89   (0.001) |
| b1 | 0.9577  (0.0085) | 0.96   (0.014) |
| b2 | -0.2546 (0.0166) | -0.25  (0.017) |
| b3 | 1.1693  (0.2123) | 2.06   (0.083) |
| b4 | 25.5313 (1.9310) | 38.67  (1.428) |
| var(a1) |  | 0.006  (0.00074) |
| var(a2) |  | 0.0001 (0.00002) |
| var(b1) |  | 0.042  (0.0042) |
| var(b2) |  | 0.063  (0.0065) |
| var(b4) |  | 52.25  (20.1) |

The mixed B-splines model was fit using the linear mixed-effects approach implemented in the TapeR package and reached convergence. In order to constrain the model predictions to equal zero at the tree top, $d_1$ which is the number of fixed parameters was set to 5 while $d_2$ which corresponds to the number of random parameters was set to 4, omitting the top most spline, respectively (see Figure A.1). The results of B-splines model fitting are represented in the following (Table 3).

Table 3: Estimates of parameters (and standard errors) for B-splines model, based on the validation data set.

| Parameters | B-splines model |
|---|---|
| $\beta_1$ | 59.1   (0.89) |
| $\beta_2$ | 41.58  (0.52) |
| $\beta_3$ | 31.76  (0.22) |

| | | |
|---|---|---|
| $\beta_4$ | 29.27 | (0.44) |
| $\beta_5$ | 13.88 | (0.05) |
| $var(\theta_1)$ | 172.95 | (13.15) |
| $var(\theta_2)$ | 114.94 | (10.72) |
| $var(\theta_3)$ | 38.71 | (6.22) |
| $var(\theta_4)$ | 136.16 | (11.66) |

Regarding the RF model, after rigorous fine-tuning, the model reached a state of convergence as the number of decision trees stabilized at an optimal value of 267. In addition, the hyperparameter tuning process led to the following optimized configuration: number of features to best split: 3 and minimum samples to split internal nodes: 4.

For each taper equation, overall statistics of fit (Bias, SEE and FI) for the entire stem were calculated and are represented in Table 4 for dob by model. Among the four models under investigation, discernible trends in bias emerged. The Max and Burkhart models exhibited a noteworthy positive bias, suggestive of their propensity to overestimate diameters. In contrast, the B-spline and Random Forest models displayed a discernible negative bias, indicating a tendency to underestimate diameters. Notably, the Max and Burkhart fixed effect model demonstrated a particularly pronounced inclination towards overestimation. The results suggest that mixed-effect models outperformed models considering only fixed effects. In particular, the Max and Burkhart and B-splines mixed-effects models showed relatively comparable SEE values of 0.78 cm and 0.96 cm, respectively. In contrast, the Max and Burkhart fixed-effect and random forest models had comparable SEE values of 1.90 cm and 2.26 cm, respectively. This comparison holds true for FI as well, although to a lesser extent than for SEE.

Table 4: Total stem fit statistics for the four taper models from cross-validation results. The underlined values refer to the best statistics. M&B is used here as the acronym of Max and Burkhart.

| Model | Bias (cm) | SEE (cm) | FI |
|---|---|---|---|
| M&B fixed effect | 0.274 | 1.90 | 0.97 |
| M&B mixed effects | 0.027 | **0.78** | **0.99** |
| B-splines | -0.015 | 0.96 | **0.99** |
| Random forest | **-0.010** | 2.26 | 0.96 |

To assess the performance of taper models at different positions throughout the stem, the statistics of fit, derived from the 5-fold cross-validation results, were analyzed by relative height classes. These statistics are documented in Table A.2 and graphically depicted in Figure 1. To comprehensively assess the taper of all models, the average bias, SEE, and FI were calculated for each model by relative height class along the stem. The results show that models including mixed effects outperformed those considering only fixed effect roughly along all the section heights. Noticeably, a significant increase of SEE was observed for all the models from a relative height class of (70 – 80 %) and was shown to coincide with the crown base height (CBH) for most of the trees (Figure 4).

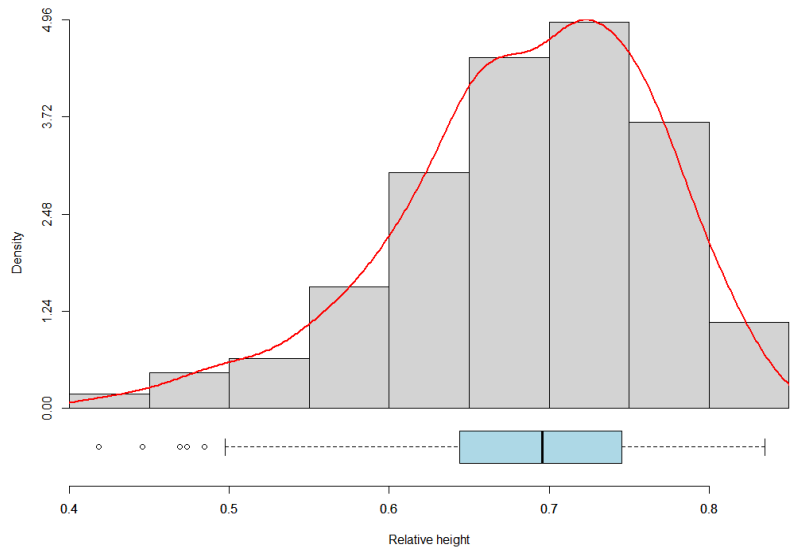

Figure 4 : Distribution of the relative crown base height (in gray color), the empirical density fit function (in red color) and, the boxplot of the relative crown base height (in blue color).

Measures of bias reveal that the Max and Burkhart fixed-effect and random forest models consistently overestimate and underestimate upper stem diameters across the entire stem, respectively. B-spline-based predictions consistently overestimate upper stem diameters in both the lower and upper stem portions while underestimating in the middle portion. In contrast, predictions from the Max and Burkhart mixed-effects models exhibit an opposing bias pattern in the lower and middle sections but demonstrate similar behavior in the upper portion of the stem.

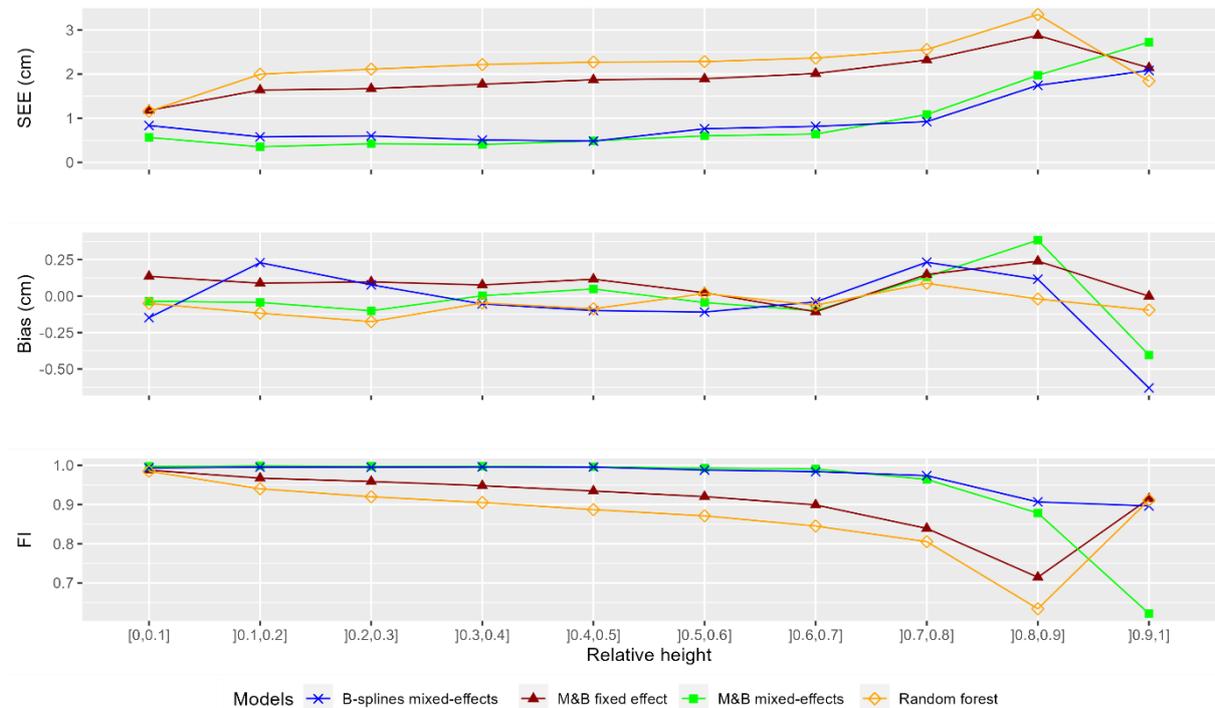

Figure 5: Bias, SEE and FI of the estimate of diameter over bark (dob) by relative class for the four models.

**Volume prediction**

The predictions of volume (over bark) by four taper models were compared with each other. Overall statistics of fit (Bias, SEE, and FI) for the total stem volume are represented in Table 5. The results show that mixed-effects models outperformed models considering only fixed effect. Specifically, Max and Burkhart and B-splines models had comparable SEE values of respectively 0.021 and 0.034 m$^3$, while the Max and Burkhart fixed effect and the random forest models had comparable SEE values of respectively 0.221 and 0.273 m$^3$. In addition, the overall results show that all the four models tend to generally overestimate the volume as the bias resulted in positive values for all the candidate models. Notably, the Max and Burkhart models (both fixed and mixed effects) exhibited the smallest overprediction, with an absolute bias value of less than 0.01 m$^3$.

Table 5: Total stem volume fit statistics for the four taper models from cross-validation results. The underlined values refer to the best statistics. M&B is used here as the acronym of Max and Burkhart.

| Model | Bias (m$^3$) | SEE (m$^3$) | FI |
|---|---|---|---|
| M&B fixed effect | 0.0085 | 0.221 | 0.97 |
| M&B mixed effects | **0.0080** | **0.021** | **0.99** |
| B-splines | 0.0240 | 0.034 | **0.99** |
| Random forest | 0.0560 | 0.273 | 0.91 |

To assess the performance of stem volume prediction for different tree dimensions, the statistics of fit (SEE, bias, and FI) were analyzed by DBH class categories. These statistics are documented in Table A.3 and graphically depicted in Figure 6. In general, results show that prediction errors are larger in large diameter trees for both Max and Burkhart fixed effect and random forest models while the prediction error tend to be constant over all the DBH classes for Max and Burkhart mixed effects and B-splines models. The results also demonstrate a strong correlation between volume predictions and volume observations (FI > ~ 0.5) for all four models across all DBH classes, except for the DBH class category (55-65 cm) and (> 65 cm), where both the M&B fixed effect and Random Forest models exhibit limitations (FI ~ 0.42 and 0.21, SEE ~ 0.39 m$^3$ and 0.71 m$^3$, respectively).

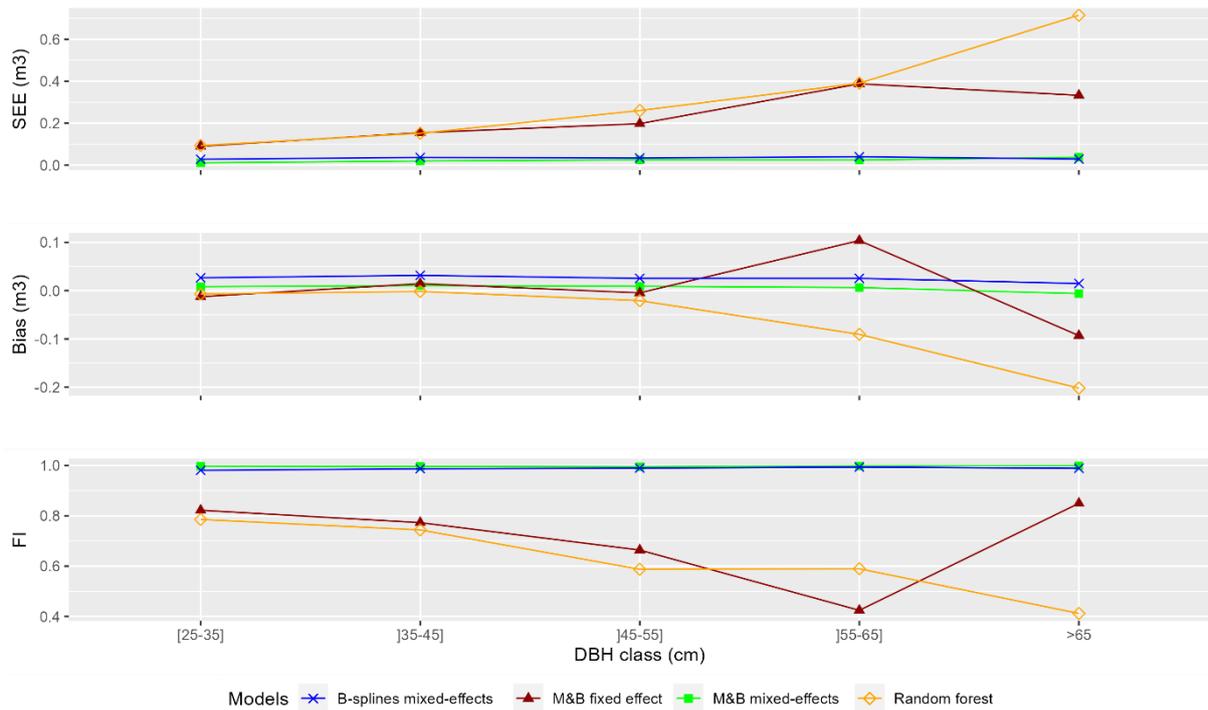

Figure 6: Bias, SEE and FI of the estimate of the stem volume by DBH class for the four models.

**Model ranking**

A taper-estimating system should provide an unbiased estimation with minimum variance of both diameter outside bark (dob) and total stem volume outside bark. Using the results from Table 4-5, it is possible to rank the four estimating systems. The rank sums were created by ranking the performance by dob and stem volume which were equally weighted here in this study. The rank sums summarized in Table 6 were generated in such a way that each estimating system was assigned a rank separately for every relative height for dob and for every DBH class for the volume. These ranks were also summed for standard errors and biases. As a result, Max and Burkhart mixed-effects system ranked first in most of all the ranking categories. B-splines based system ranked second in estimating diameter and volume by standard error of estimates while random forest system ranked third and Max and Burkhart fixed effect ranked fourth with regards to the same previously discussed estimates.

Table 6: Rank sum analysis for the four estimating systems.

| Model / Parameters | Max and Burkhart fixed effect | Max and Burkhart mixed effects | B-splines | Random forest |
|---|---|---|---|---|
| Overall SEE for diameter and volume | 6 (3) | 2 (1) | 4 (2) | 8 (4) |
| Overall bias for diameter and volume | 6 (3) | 4 (1) | 5 (2) | 5 (2) |
| SEE for diameter by relative height | 31 (3) | 17 (1) | 16 (2) | 36 (4) |
| Bias for diameter by relative height | 29 (3) | 21 (2) | 30 (4) | 20 (1) |
| SEE for volume by DBH classes | 15 (3) | 6 (1) | 8 (2) | 18 (4) |

| Bias for volume by DBH classes | 16 (4) | 9 (1) | 13 (3) | 12 (2) |
|---|---|---|---|---|
| **Total** | 103 (4) | 59 (1) | 76 (2) | 99 (3) |

## Discussion

Taper equations are an indispensable tool in the field of forest management and industry. However, the development of an accurate taper model exhibits inherent challenges, mainly due to the need for comprehensive customization to account for the characteristics exhibited by each species and site-specific environmental conditions, and the availability of high-quality data required for the model development. In the present study, four taper models from three different model categories were compared in terms of their capacity to predict diameter over bark (dob) and total stem volume of *Pinus nigra* species. Across models, the Max and Burkhart and B-splines mixed-effects models demonstrated superior accuracy and reliability in predicting both dob and total stem volume, with a small advance for the Max and Burkhart model.

The Terrestrial Laser Scanner (TLS) has previously been employed as an effective tool for measuring trees in forest environments and generating taper equations (Saarinen et al., 2019; Sun et al., 2016). The case study conducted on black pine forests, which is presented here, demonstrated the feasibility of using TLS for surveys and highlighted the high quality of the acquired data. In the Italian context, where tree measurements for quantitative purposes have significantly decreased in recent decades, technological tools like TLS represent a pivotal advancement for rejuvenating our understanding of the Italian forest sector.

In the present study, biases and standard errors of estimate were evaluated for diameter outside bark (dob) and total stem volume. These biases and standard errors were also evaluated for different heights within the trees and across different tree sizes. The overall goodness-of-fit statistics (B, SEE, and FI) were calculated and reported in Table 4-5 and Figure 2-3 for the four tested models. The results indicate that the Max and Burkhart and B-splines mixed-effects models explained more than 99% of the total variation in predicting the upper-stem diameter and total stem volume. The Max and Burkhart fixed-effect model explained more than 96% of this total variation, while the random forest model explained more than 96% for the diameter variation and 91% for the total volume variation.

The superiority of models that include random effects over those that only consider fixed effects has been demonstrated in various previous studies, and our findings are in agreement with this. For example, a study conducted by (Leites & Robinson, 2004) showed that the inclusion of random effects considerably improved the fit of the Max and Burkhart taper equation for the *Pinus taeda* species. In a different study by (Bronisz & Zasada, 2020), the Kozak's taper equation fitted using a mixed-effects approach provided better results for the upper diameter and total stem volume for the *Pinus sylvestris* species. In our study, the Max & Burkhart (1976) model was fitted using both fixed and mixed effects approaches. Notably, the error was reduced by 52% for the upper-stem diameter and by 89% for the total stem volume when random effects were additionally considered.

It is also noteworthy that the evaluation of the proposed models exhibited a higher Standard Error of Estimation (SEE) in predicting the upper-stem diameter at a relative height of [0.7, 0.8] (Table A.2). The analysis of the distribution of the tree's relative crown base height reveals that 50% of the observations are located between a value of 0.7 and 0.8, and the maximum relative crown base height was found to be approximately equal to 0.8 (Figure 4). This implies that all observations above a relative height of 0.8 were necessarily measured inside the crown. This last result may partly explain the notable increase in error in predicting the upper-stem diameter at a relative height of 0.7. In fact, many studies have reported the inability of Terrestrial Laser Scanners to collect reliable data beyond the base of the crown due to interference from branches and leaves. Additionally, as the distance from the scanner increases, spatial resolution decreases, which is a well-known effect in scanning. This reduction in resolution can substantially impact the upper parts of trees, in particular (Liang et al., 2016). The interference from the crown, compounded with the increased distance, may explain the models' inability to detect a well-defined taper pattern above the crown base.

In this study, Monte Carlo numerical integration was employed for estimating the volume of tree stems using the non-differentiable and complex Random Forest model. This decision was justified by the model's inherent characteristics mentioned earlier, which render traditional integration methods less suitable. It is imperative to underscore the significance of comparing our chosen method with alternative numerical integration techniques to gain insights into result robustness and accuracy, potentially informing future research. Notably, the RF model, unlike the parametric and semi-parametric models used in this study, typically demands a substantial amount of training data. In our analysis, the RF model obtained the lowest score in terms of SEE (dob and volume). This highlights the potential impact of our relatively small dataset (219 trees) on RF performance. In addition, the use of various integration methods (Huber, Monte Carlo, and analytical integration) in our study may introduce minor biases affecting our results. Further research is needed to assess and potentially mitigate theses biases, as their combined impact on the findings remains uncertain.

Considering the results of this study, the Max & Burkhart (1976) taper model, calibrated by means of a mixed-effects approach, and its compatible volume integration could be considered for operational use in estimating diameter outside bark at different heights, as well as total stem volume, respectively, with an overall precision of 0.781 cm and 0.021 m³ for the *Pinus nigra* species.

## Conclusion

In this study, four taper models from three different model classes were evaluated in terms of their capacity to predict diameter at different heights and the total stem volume. The data used in this research was collected non-destructively using TLS technology. The Max & Burkhart (1976) taper model calibrated with a mixed-effects approach was superior in estimating diameter and volume in comparison with all the other candidate models. Nevertheless, the imprecise diameter measurements deduced from TLS inside crown prohibits a rigorous evaluation of less valuable parts of the stem for all models. The selected developed model within the frame of this study can be operationally used by forest managers to better predict diameter and stem volume as well as to help optimize the allocation of wood to different harvested products.

**Appendix A**

Table A.1: Model selection statistics for evaluating the inclusion of random parameters in the Max and Burkhart taper model.

| Combination | AIC | BIC | Combination | AIC | BIC |
|---|---|---|---|---|---|
| b4 | -15057,955 | -15030,843 | a1-b4 | -18626,903 | -18596,402 |
| b3 | -15057,955 | -15030,843 | a1-b3 | -18851,642 | -18821,14 |
| b3-b4 | -15055,847 | -15025,345 | a1-b3-b4 | -18838,969 | -18805,078 |
| b2 | -21220,618 | -21193,506 | a1-b2 | -22732,771 | -22702,269 |
| b2-b4 | -21220,31 | -21189,809 | a1-b2-b4 | -22715,597 | -22681,706 |
| b2-b3 | -22291,665 | -22261,164 | a1-b2-b3 | -22749,4 | -22715,509 |
| b2-b3-b4 | -15885,951 | -15852,06 | a1-b2-b3-b4 | -22759,431 | -22722,152 |
| b1 | -22550,526 | -22523,413 | a1-b1 | -24104,749 | -24074,247 |
| b1-b4 | -24435,313 | -24404,812 | a1-b1-b4 | -24044,618 | -24010,727 |
| b1-b3 | -16443,308 | -16412,807 | a1-b1-b3 | -24439,643 | -24405,752 |
| b1-b3-b4 | -16441,076 | -16407,185 | a1-b1-b3-b4 | -24436,415 | -24399,135 |
| b1-b2 | -16449,376 | -16418,875 | a1-b1-b2 | -25954,858 | -25920,968 |
| b1-b2-b4 | -16448,498 | -16414,608 | a1-b1-b2-b4 | -25971,313 | -25934,034 |
| b1-b2-b3 | -16448,208 | -16414,318 | a1-b1-b2-b3 | -26134,871 | -26097,591 |
| b1-b2-b3-b4 | -16445,24 | -16407,96 | a1-b1-b2-b3-b4 | -26132,522 | -26091,853 |
| a2 | -18850,569 | -18823,457 | a1-a2 | -22457,125 | -22426,624 |
| a2-b4 | -18723,922 | -18693,421 | a1-a2-b4 | -22425,911 | -22392,02 |
| a2-b3 | -18848,253 | -18817,752 | a1-a2-b3 | -18791,895 | -18758,004 |
| a2-b3-b4 | -18681,531 | -18647,641 | a1-a2-b3-b4 | -18839,839 | -18802,559 |
| a2-b2 | -22972,719 | -22942,217 | a1-a2-b2 | -24133,78 | -24099,889 |
| a2-b2-b4 | -23565,037 | -23531,146 | a1-a2-b2-b4 | -24138,454 | -24101,174 |
| a2-b2-b3 | -22773,131 | -22739,24 | a1-a2-b2-b3 | -23371,059 | -23333,779 |
| a2-b2-b3-b4 | -22699,864 | -22662,585 | a1-a2-b2-b3-b4 | -22763,72 | -22723,051 |
| a2-b1 | -24416,318 | -24385,817 | a1-a2-b1 | -25843,329 | -25809,438 |
| a2-b1-b4 | -24389,32 | -24355,429 | a1-a2-b1-b4 | -25616,892 | -25579,613 |
| a2-b1-b3 | -24408,473 | -24374,582 | a1-a2-b1-b3 | -24962,533 | -24925,253 |
| a2-b1-b3-b4 | -24422,755 | -24385,475 | a1-a2-b1-b3-b4 | -24428,357 | -24387,688 |
| a2-b1-b2 | -26661,474 | -26627,584 | a1-a2-b1-b2 | -27212,711 | -27175,431 |
| a2-b1-b2-b4 | -26144,912 | -26107,633 | **a1-a2-b1-b2-b4** | **-27249,686** | **-27209,017** |
| a2-b1-b2-b3 | -26136,166 | -26098,886 | a1-a2-b1-b2-b3 | -26857,154 | -26816,485 |
| a2-b1-b2-b3-b4 | -26156,818 | -26116,149 | a1-a2-b1-b2-b3-b4 | -26890,248 | -26846,19 |
| a1 | -18697,024 | -18669,911 | | | |

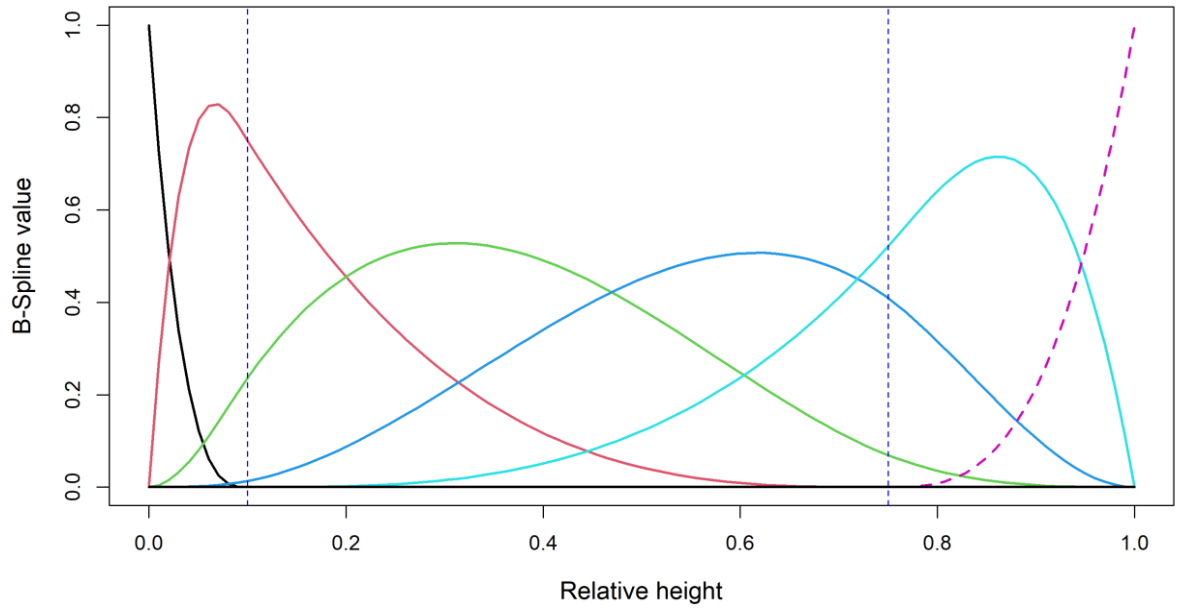

Figure A.1: Population mean-fixed effects B-spline basis, l = 1, 2, 3, 4, 5, 6. Dashed line indicate the omitted spline in order to constrain the equation for the total height. Blue vertical lines correspond to the knot's values. The same omission is applied in the B-spline basis $B^{(2)}(hr)$ for the tree-specific deviation from the mean response.

1 Table A.2: Bias (cm), standard error (cm), and fitness index of the estimate of dob by relative height for four models.

| Relative height | No bs | M&B fixed effect | | | M&B mixed-effects | | | B-splines | | | RF | | |
|---|---|---|---|---|---|---|---|---|---|---|---|---|---|
| | | SEE ± SE | Bias ± SE | FI ± SE | SEE ± SE | Bias ± SE | FI ± SE | SEE ± SE | Bias ± SE | FI ± SE | SEE ± SE | Bias ± SE | FI ± SE |
| 0-0.1 | 745 | 1,177 ± 0,116 | 0,135 ± 0,149 | 0,987 ± 0,004 | 0,563 ± 0,054 | -0,035 ± 0,015 | 0,997 ± 0,001 | 0,834 ± 0,019 | -0,148 ± 0,047 | 0,993 ± 0,002 | 1,157 ± 0,294 | -0,050 ± 0,173 | 0,985 ± 0,008 |
| 0.1-0.2 | 654 | 1,639 ± 0,252 | 0,089 ± 0,236 | 0,967 ± 0,012 | 0,352 ± 0,020 | -0,043 ± 0,019 | 0,999 ± 0,001 | 0,578 ± 0,065 | 0,229 ± 0,060 | 0,995 ± 0,002 | 1,996 ± 0,231 | -0,117 ± 0,312 | 0,940 ± 0,025 |
| 0.2-0.3 | 655 | 1,670 ± 0,158 | 0,098 ± 0,196 | 0,959 ± 0,013 | 0,421 ± 0,026 | -0,100 ± 0,048 | 0,998 ± 0,001 | 0,598 ± 0,059 | 0,077 ± 0,062 | 0,995 ± 0,001 | 2,113 ± 0,257 | -0,175 ± 0,338 | 0,920 ± 0,040 |
| 0.3-0.4 | 676 | 1,772 ± 0,155 | 0,076 ± 0,177 | 0,948 ± 0,016 | 0,403 ± 0,020 | 0,003 ± 0,013 | 0,997 ± 0,001 | 0,507 ± 0,040 | -0,054 ± 0,026 | 0,996 ± 0,001 | 2,216 ± 0,284 | -0,048 ± 0,462 | 0,905 ± 0,045 |
| 0.4-0.5 | 645 | 1,870 ± 0,195 | 0,116 ± 0,190 | 0,934 ± 0,024 | 0,486 ± 0,051 | 0,049 ± 0,055 | 0,996 ± 0,001 | 0,481 ± 0,056 | -0,099 ± 0,051 | 0,995 ± 0,001 | 2,273 ± 0,271 | -0,087 ± 0,423 | 0,887 ± 0,052 |
| 0.5-0.6 | 652 | 1,894 ± 0,240 | 0,023 ± 0,139 | 0,920 ± 0,032 | 0,600 ± 0,048 | -0,044 ± 0,055 | 0,993 ± 0,002 | 0,761 ± 0,081 | -0,110 ± 0,069 | 0,988 ± 0,003 | 2,283 ± 0,260 | 0,017 ± 0,334 | 0,871 ± 0,068 |
| 0.6-0.7 | 644 | 2,013 ± 0,220 | -0,107 ± 0,171 | 0,899 ± 0,041 | 0,640 ± 0,049 | -0,100 ± 0,047 | 0,990 ± 0,003 | 0,816 ± 0,101 | -0,040 ± 0,080 | 0,984 ± 0,005 | 2,363 ± 0,252 | -0,062 ± 0,336 | 0,845 ± 0,080 |
| 0.7-0.8 | 660 | 2,320 ± 0,205 | 0,148 ± 0,214 | 0,839 ± 0,067 | 1,084 ± 0,252 | 0,133 ± 0,101 | 0,964 ± 0,027 | 0,922 ± 0,195 | 0,232 ± 0,091 | 0,973 ± 0,019 | 2,556 ± 0,349 | 0,087 ± 0,468 | 0,805 ± 0,077 |
| 0.8-0.9 | 547 | 2,876 ± 0,225 | 0,240 ± 0,211 | 0,714 ± 0,082 | 1,974 ± 0,190 | 0,384 ± 0,235 | 0,878 ± 0,027 | 1,745 ± 0,200 | 0,115 ± 0,225 | 0,906 ± 0,015 | 3,351 ± 0,391 | -0,020 ± 0,422 | 0,634 ± 0,043 |
| 0.9-1 | 316 | 2,143 ± 0,569 | -0,001 ± 0,501 | 0,915 ± 0,010 | 2,724 ± 0,431 | -0,405 ± 0,803 | 0,622 ± 0,301 | 2,082 ± 0,388 | -0,630 ± 0,206 | 0,896 ± 0,038 | 1,842 ± 0,295 | -0,096 ± 0,316 | 0,911 ± 0,023 |

Table A.3: Bias (m3), standard error (m3), and fitness index of the estimate of stem volume by DBH class for four models

| DBH class (cm) | Nobs | M&B fixed effect | | | M&B mixed-effects | | | B-splines | | | RF | | |
|---|---|---|---|---|---|---|---|---|---|---|---|---|---|
| | | SEE ± SE | Bias ± SE | FI ± SE | SEE ± SE | Bias ± SE | FI ± SE | SEE ± SE | Bias ± SE | FI ± SE | SEE ± SE | Bias ± SE | FI ± SE |
| 25-35 | 35 | 0,090 ± 0,027 | -0,013 ± 0,023 | 0,822 ± 0,106 | 0,011 ± 0,001 | 0,008 ± 0,001 | 0,997 ± 0,001 | 0,028 ± 0,003 | 0,027 ± 0,002 | 0,981 ± 0,011 | 0,093 ± 0,024 | -0,007 ± 0,029 | 0,786 ± 0,104 |
| 35-45 | 83 | 0,155 ± 0,055 | 0,014 ± 0,055 | 0,773 ± 0,132 | 0,020 ± 0,002 | 0,010 ± 0,001 | 0,997 ± 0,001 | 0,037 ± 0,003 | 0,032 ± 0,003 | 0,987 ± 0,003 | 0,152 ± 0,022 | -0,002 ± 0,027 | 0,744 ± 0,171 |
| 45-55 | 56 | 0,198 ± 0,029 | -0,005 ± 0,051 | 0,664 ± 0,182 | 0,025 ± 0,002 | 0,009 ± 0,001 | 0,995 ± 0,001 | 0,034 ± 0,007 | 0,025 ± 0,010 | 0,990 ± 0,005 | 0,260 ± 0,030 | -0,021 ± 0,071 | 0,588 ± 0,125 |
| 55-65 | 33 | 0,388 ± 0,052 | 0,104 ± 0,109 | 0,424 ± 0,489 | 0,024 ± 0,003 | 0,006 ± 0,002 | 0,998 ± 0 | 0,040 ± 0,008 | 0,025 ± 0,010 | 0,994 ± 0,004 | 0,391 ± 0,160 | -0,091 ± 0,032 | 0,590 ± 0,281 |
| > 65 | 10 | 0,333 ± 0,071 | -0,093 ± 0,245 | 0,850 ± 0,027 | 0,038 ± 0,002 | -0,006 ± 0,007 | 0,999 ± 0,001 | 0,029 ± 0,019 | 0,015 ± 0,017 | 0,989 ± 0,015 | 0,715 ± 0,470 | -0,202 ± 0,604 | 0,412 ± 0,258 |